\documentclass[graybox]{svmult}
\usepackage[numbers,comma,sort&compress]{natbib}

\usepackage{bookmark}
\makeatletter
\providecommand*{\toclevel@titlech}{0}
\edef\toclevel@authorch{1000}
\makeatother

\usepackage{mathptmx}       
\usepackage{bm}
\usepackage{helvet}         
\usepackage{type1cm}        
\usepackage{makeidx}         
\usepackage{graphicx}        
\usepackage{multicol}        
\usepackage[bottom]{footmisc}
\usepackage[english]{babel}
\usepackage{textcomp}
\usepackage{url}

\usepackage{color}
\definecolor{darkblue}{rgb}{0,0,.6}
\usepackage{hyperref}
\hypersetup{
  bookmarksnumbered = true,
  colorlinks = true, linkcolor = darkblue,
  citecolor = darkblue, filecolor = darkblue,
  menucolor = darkblue, urlcolor = darkblue
}

\makeindex

\usepackage{amsmath,amssymb}

\newcommand{\iu}{\mathrm{i}}	
\newcommand{\de}{\mathrm{d}}	
\newcommand{\ee}{\mathrm{e}}	

\newcommand{\melement}[3]{\ensuremath{\left\langle #1\left| #2 \right| #3\right\rangle}}

\newcommand\Berry{\ensuremath{\bm{\xi}}}
\newcommand\En{\ensuremath{E}}

\begin{document}

\title{Ultrafast control of strong-field electron dynamics in solids}
\author{%
Vladislav S.~Yakovlev$^{1,2}$,
Stanislav Yu.~Kruchinin$^1$,
Tim Paasch--Colberg$^1$,\\
Mark I.~Stockman$^3$,
Ferenc Krausz$^{1,2}$}

\authorrunning{Vladislav S.~Yakovlev, Stanislav Yu.~Kruchinin, Tim Paasch--Colberg \emph{et al.}}

\institute{%
$^1$ Max-Planck-Institut f\"ur Quantenoptik, Hans-Kopfermann-Stra{\ss}e 1, 85748 Garching, Germany,
\email{vladislav.yakovlev@mpq.mpg.de}
\\
$^2$ Ludwig-Maximilians-Universit\"at, Am Coulombwall~1, 85748 Garching, Germany
\\
$^3$ Center for Nano-Optics (CeNO) and Department of Physics and Astronomy, Georgia State University, Atlanta, Georgia 30340, USA}

\maketitle

\abstract{We review theoretical foundations and some recent progress related to
  the quest of controlling the motion of charge carriers with intense laser
  pulses and optical waveforms. The tools and techniques of attosecond science
  enable detailed investigations of a relatively unexplored regime of
  nondestructive strong-field effects. Such extremely nonlinear effects may be
  utilized to steer electron motion with precisely controlled optical fields and
  switch electric currents at a rate that is far beyond the capabilities of
  conventional electronics.}

\noindent

\section{Introduction}
\label{sec:introduction}

It has long been realized that intense few-cycle laser pulses provide unique
conditions for exploring extremely nonlinear phenomena in
solids~\cite{Lenzner_1998_PRL_80_4076,Mao_APA_2004_79_1695}, the key idea being
that a sample can withstand a stronger electric field if the duration of the
interaction is shortened. Ultimately, a single-cycle laser pulse provides the
best conditions for studying nonperturbative strong-field effects, especially
those where the properties of a sample change within a fraction of a laser
cycle. The recent rapid development of the tools and techniques of attosecond
science~\cite{Krausz_2009_RMP_81_163} not only creates new opportunities for
detailed investigations of ultrafast electron dynamics in solids, but it also
opens exciting opportunities for controlling electron motion in solids with
unprecedented speed and accuracy. Conventional nonlinear phenomena that
accompany the interaction of intense laser pulses with solids have already found
a vast number of applications in spectroscopy, imaging, laser technology,
transmitting and processing information~\cite{Garmire_OE_2013_21_30532}. It can
be expected that the less conventional nonperturbative nonlinearities may also
find important applications~\cite{Krausz_NaturePhotonics_2014,Ghimire_JPB_2014}.
The main purpose of this chapter is to review
theoretical foundations and some recent progress in this field.

Most of the relevant physical phenomena are well known, such as interband
tunneling, Franz--Keldysh effect, Bloch oscillations, and Wannier--Stark
localization. However, new experiments put these phenomena in a new context,
which often leads to nontrivial observations, such as the generation of
nonperturbative high-order harmonics in a solid due to Bragg-like scattering at
the edges of the Brillouin zone
\cite{Ghimire_2011_NP_7_138,Schubert_2014_NP_8_119}, a nearly instantaneous
change in extreme-ultraviolet absorptivity and near-infrared reflectivity of a
dielectric in the presence of a laser field as strong as several volts per
{\aa}ngstr\"om \cite{Schultze_2013_Nature_493_75}, or the induction of electric
current in an unbiased dielectric by similarly intense laser pulses
\cite{Schiffrin_2013_Nature_493_70}. Several decades of research on strong-field
phenomena in solids and mesoscopic structures provide a solid ground for
developing new theoretical models adapted for new experimental conditions. At
the same time, a description of extremely nonlinear processes that unfold during just
a few femtoseconds requires approximations that may be different from those
established for longer and less intense laser pulses. Both \textit{ab initio}
calculations that attempt to capture all the complexity of many-electron
dynamics and ``toy models'' designed to deepen our understanding of basic
phenomena are going to play an important role in extending our ability to
control the optical and electric properties of solids with controlled light
fields.

The key parameters that determine the regime of strong-field light--matter
interaction with dielectrics and semiconductors are the Keldysh parameter
\begin{equation}
  \label{eq:Keldysh_gamma}
  \gamma_\mathrm{K} = \frac{\omega_\mathrm{L} \sqrt{m E_g}}{e F_\mathrm{L}},
\end{equation}
the Bloch frequency
\begin{equation}
  \label{eq:Bloch_frequency}
  \omega_\mathrm{B} = \frac{e F_\mathrm{L} a}{\hbar},
\end{equation}
and the Rabi frequency
\begin{equation}
  \label{eq:Rabi_frequency}
  \Omega_\mathrm{R} = \frac{d_\mathrm{cv} F_\mathrm{L}}{\hbar}.
\end{equation}
Here, $F_\mathrm{L}$ is the amplitude of a linearly polarized electric field
oscillating at an angular frequency $\omega_\mathrm{L}$, $e>0$ is the absolute
value of the electron charge, $m$ is the reduced mass of an electron and a hole
($m^{-1} = m_\mathrm{e}^{-1} + m_\mathrm{h}^{-1}$), $a$ is the lattice period in
the polarization direction of the electric field, and $d_\mathrm{cv}$ is the
dipole matrix element responsible for transitions between valence- and
conduction-band states. Conditions that have to be fulfilled for perturbation
theory to be applicable include $\gamma_\mathrm{K} \gg 1$, $\omega_\mathrm{B}
\ll \omega_\mathrm{L}$, $\Omega_\mathrm{R} \ll \omega_\mathrm{L}$, and
$\hbar \Omega_\mathrm{R} \ll E_\mathrm{g}$. Correspondingly, the main physical effects
that make the interaction nonperturbative are interband tunneling
($\gamma_\mathrm{K} \lesssim 1$), Bloch oscillations ($\omega_\mathrm{B} \ge \pi
\omega_\mathrm{L}$),
Rabi flopping ($\hbar \Omega_\mathrm{R} \gtrsim E_\mathrm{g}$),
and carrier--wave Rabi flopping ($\Omega_\mathrm{R} \gtrsim
\omega_\mathrm{L}$), the last two effects being particularly important
for resonant excitations ($\hbar \omega_\mathrm{L} \approx E_\mathrm{g}$).
For a band gap of several electronvolts and a laser
frequency in the near-infrared spectral range ($\lambda_\mathrm{L} = 2 \pi c /
\omega_\mathrm{L} \sim 1$~\textmu{}m), these effects become essential for
$F_\mathrm{L} \gtrsim 1$~V/\AA{}, which corresponds to a laser
intensity of $I_\mathrm{L} \gtrsim 10^{13}\ \mbox{W}/\mbox{cm}^2$~\cite{Wegener_2005}.

\section{Main theoretical concepts}\label{sec:basic_theory}

For a theoretical description of phenomena that take place on a few-femtosecond
time scale, it is common (although not necessarily correct) to neglect dephasing.
In this case, the time-dependent Schr\"odinger equation (TDSE)
\begin{equation}
  \label{eq:TDSE}
  \iu \hbar \partial_t \psi(t) = \hat{H}(t) \psi(t)
\end{equation}
fully describes nonrelativistic electron dynamics. In situations where
phase and energy relaxation processes are important, one has to use the more general
formalisms of density matrices or nonequilibrium Green's functions, which are
beyond the scope of this chapter. In the following, we review some theoretical
concepts developed for the case of a single charged particle moving in a
periodic potential. Furthermore, since the wavelength of visible or infrared
light is much larger than the size of a unit cell, and electron velocities are
much smaller than the speed of light, we can use the dipole approximation, which
neglects the spatial dependence of the laser field while solving the
TDSE: $\mathbf{F}_\mathrm{L} = \mathbf{F}_\mathrm{L}(t)$.

The first step to solve the TDSE is to choose a gauge and a basis. The exact
solution does not depend on this choice, but the chosen gauge and basis dictate
approximations that one may wish to make, and they influence the physical
interpretation of results. In the dipole approximation, the two main options
are the velocity and length gauges, which, in the following, will be
abbreviated as ``VG'' and ``LG'', respectively. These two gauges are related to
each other by the following unitary transformation of the respective wave
functions $\psi_\mathrm{VG}$ and $\psi_\mathrm{LG}$:
\begin{equation}
  \label{eq:LG_to_VG}
  \psi_\mathrm{VG}(t) = \exp\left[-\frac{\iu}{\hbar} e \mathbf{A}_\mathrm{L}(t) \mathbf{r}\right]
    \psi_\mathrm{LG}(t),
\end{equation}
where
\begin{equation}
  \label{eq:vector_potential}
  \mathbf{A}_\mathrm{L}(t) = -\int^t \mathbf{F}_\mathrm{L}(t')\,\de t'
\end{equation}
is the vector potential of the laser field.

The Hamilton operators in these two gauges take the following forms:
\begin{align}
  \label{eq:H_VG}
  \hat{H}_\mathrm{VG} &= \frac{\left[\hat{\mathbf{p}} + e \mathbf{A}_\mathrm{L}(t)\right]^2}{2 m} + U(\mathbf{r}),\\
  \label{eq:H_LG}
  \hat{H}_\mathrm{LG} &= \frac{\hat{\mathbf{p}}^2}{2 m} + U(\mathbf{r}) + e \mathbf{F}_\mathrm{L}(t) \mathbf{r}.
\end{align}
A big advantage of the velocity gauge for numerical simulations is that a
homogeneous external field does not destroy the spatial periodicity of the
Hamiltonian, so that the Bloch theorem applies even in the presence of the
field. At the same time, transformation \eqref{eq:LG_to_VG} can be interpreted
as a transition to a moving coordinate system, in which electrons acquire an
additional momentum $-e \mathbf{A}_\mathrm{L}(t)$. In order to accurately
account for such superficial dynamics in a time-independent basis, a sufficient
number of basis states and exact transition matrix elements are required.
Calculations in the velocity gauge are particularly problematic in the limit of
a static field ($\omega_\mathrm{L} \to 0$).

These problems are circumvented in the length gauge. The price for this is the
fact that the term $e \mathbf{F}_\mathrm{L}(t) \mathbf{r}$, which is responsible
for the interaction with an external field, destroys the spatial periodicity of
the Hamiltonian. Also, when periodic boundary conditions are applied in the
length gauge, the interaction potential becomes discontinuous at the boundaries
of the unit cell. Nevertheless, once these difficulties are addressed, the
length gauge becomes an appropriate choice for numerical simulations
\cite{Aversa_1995_PRB_52_14636,Resta_1998_PRL_80_1800,Souza_2004_PRB_69_085106,
  Virk_2007_PRB_76_035213,Springborg_2008_PRB_77_045102}.

When the field of a laser pulse is nonresonant ($\hbar \omega_\mathrm{L} \ll
E_\mathrm{g}$) and strong, it is frequently convenient to use a time-dependent
basis of quantum states that adiabatically ``adapt'' themselves to the external
field. In the length gauge, such an adiabatic basis is given by Wannier--Stark
states. In the velocity gauge, this role is played by accelerated Bloch states,
also known as Houston functions~\cite{Houston_1940_PR_57_184}. The following two
subsections summarize the most important properties of these states.

\subsection{Wannier--Stark resonances}
\label{sec:Wannier}

For simplicity, let us consider a one-dimensional problem. For a constant
external field parallel to the $z$-axis, the stationary Schr\"odinger equation reads
\begin{equation}
  \label{eq:WS_problem}
  \hat{H}_\mathrm{LG} \psi \equiv
  \left(\frac{\hat{p}^2}{2 m} + U(z) + e F_\mathrm{L} z\right) \psi =
  \En \psi.
\end{equation}
The potential here is periodic: $U(z+a) = U(z)$ with a lattice period $a$.
Leaving the question of the existence of such eigenstates aside, one can ask
which properties the solutions of \eqref{eq:WS_problem} possess if they exist.
From the periodicity of the
potential, we immediately conclude that if $\psi(z)$ is an eigenstate with an
energy $\En$, then $\psi(z-a)$ is also an eigenstate with the energy $\En + e
F_\mathrm{L} a$. The additional term $e F_\mathrm{L} a$ is the energy required
to move an electron against the laser field by one lattice period. This suggests
that the eigenstates $\psi(z)$ should be localized functions. They are referred
to as Wannier--Stark states.

Wannier found approximate solutions to \eqref{eq:WS_problem}
by defining an auxiliary problem
\begin{equation}
  \left( \frac{\hat{p}^2}{2 m} + U(z) + e F_\mathrm{L}
    \left[ z + \iu \frac{\partial}{\partial k} \right] \right) b_i(z,k) = \En_i(k) b_i(z,k),
\end{equation}
where $i$ is a band index, $k$ is the crystal momentum, and $\En_i(k)$ is the
energy of the unperturbed Bloch state. Wannier showed
\cite{Wannier_1960_PR_117_432} that the solutions of this problem, known as
Wannier--Bloch states, decouple different bands. Specifically, knowing
$b_i(z,k)$ that satisfies the periodic boundary condition in reciprocal
space $b_i(z,k+2\pi/a) = b_i(z,k)$, an approximate solution of the
time-dependent Schr\"odinger equation with the Hamiltonian $H_\mathrm{LG}$ can
be written as
\begin{equation}
  \psi_i(z,t) = b_i\left(z, k_0-\frac{e F_\mathrm{L}}{\hbar} t \right)
  \exp\left[ -\frac{\iu}{\hbar} \int_{t_0}^t \En_i\left(k_0-\frac{e F_\mathrm{L}}{\hbar} t' \right)\,\de t' \right],
\end{equation}
where the approximation consists in restricting the electron motion to a single
band, that is, neglecting interband transitions. Obviously, $|\psi_i(z,t)|^2$ is
a periodic function of time, the period being equal to the period of Bloch
oscillations: $T_\mathrm{B} = 2 \pi / \omega_\mathrm{B} = 2 \pi \hbar / (e
F_\mathrm{L} a)$. The Wannier--Stark states for a bulk crystal
$\psi_{i,\ell}^\mathrm{WS}(z)$ are defined via the expansion
\begin{equation}
  \psi_i(z,t) = \sum_\ell \psi_{i,\ell}^\mathrm{WS}(z) \exp\left[-\frac{\iu}{\hbar} \En_{i,\ell}^\mathrm{WS} t\right].
\end{equation}
A state $\psi_{i,\ell}^\mathrm{WS}(z)$ is localized at a lattice site $\ell$.
Explicit expressions for the Wannier--Stark states and their energies are
\begin{gather}
  \psi_{i,\ell}^\mathrm{WS}(z) = \frac{a}{2 \pi} \int_{-\pi/a}^{\pi/a} \de k\, b_i(z, k) \ee^{-\iu \ell a k},\\
  \label{eq:WS_energies}
  \En_{i,\ell}^\mathrm{WS} = \overline{\En}_i + \ell e a F_\mathrm{L},
\end{gather}
where $\overline{\En}_i$ is the mean energy of band $i$:
\begin{equation*}
  \overline{\En}_i = \frac{a}{2 \pi} \int_{-\pi/a}^{\pi/a} \de k\, \En_i(k).
\end{equation*}

The energies $\En_{i,\ell}^\mathrm{WS}$ form the so-called ``Wannier--Stark
ladder''---plotted against $F_\mathrm{L}$, they are a set of straight lines,
where the slope of each line $\de \En_{i,\ell}^\mathrm{WS}/\de F_\mathrm{L} = \ell e a$ is
determined by the lattice index $\ell$.


The localization length of a Wannier--Stark state is given by
\begin{equation}
  L_i^\mathrm{WS} = \frac{\Delta_i}{e |F_\mathrm{L}|},
\end{equation}
where $\Delta_i$ is the energy interval covered by band $i$.

Our introduction to the Wannier--Stark states has so far followed the one given
by Wannier~\cite{Wannier_1962_RMP_34_645}. The existence of Wannier--Stark
states had been a subject of numerous disputes for three decades until they were
experimentally observed in superlattices~\cite{Voisin_1988_PRL_61_1639}. The
core of these disputes was the question whether the Wannier--Stark states retain
their physical significance when interband transitions are accounted
for~\cite{Zak_1968_PRL_20_1477,Wannier_1969_PR_181_1364,Avron_1977_JMP_18_918}.
According to modern
treatments~\cite{Gluck_2002_PR_366_103,Resta_2000_JPCM_12_R107,%
  Sundaram_1999_PRB_59_14915,Xiao_2010_RMP_82_1959}, Wannier--Stark states
should be viewed as resonances (metastable states) with lifetimes
$\tau_i^\mathrm{WS} = 1/\Gamma_i$, and \eqref{eq:WS_energies} should be
generalized as
\begin{equation}
  \En_{i,\ell}^\mathrm{WS} = \overline{\En}_i + 
  \left(\ell - \frac{\gamma_i^\mathrm{Zak}}{2\pi} \right) e a F_\mathrm{L}
  - \iu \hbar \frac{\Gamma_i}{2}.
\end{equation}
Here, $\Gamma_i$ is the decay rate due to interband transitions, and
\begin{equation}
  \label{eq:Zak_phase}
  \gamma_i^\mathrm{Zak} = \oint_\mathrm{BZ} \de\mathbf{k}\cdot\Berry_{ii}(\mathbf{k})
\end{equation}
is called Zak's phase~\cite{Resta_2000_JPCM_12_R107,Zak_1989_PRL_62_2747}, where
the integral is taken over a smooth closed path across the entire Brillouin
zone, and $\Berry_{ii}(\mathbf{k})$ is the Berry connection or geometric
vector
potential~\cite{Resta_2000_JPCM_12_R107,Sundaram_1999_PRB_59_14915,Xiao_2010_RMP_82_1959}.
An explicit expression for $\Berry_{ij}(\mathbf{k})$ is given by
\eqref{eq:Blount} in the next section. Zak's phase plays an important role in
the ``modern theory of polarization''~\cite{King-Smith_PRB_1993}; it is equal to
either 0 or $\pi$ for crystals that possess inversion symmetry, and it can
assume any value for other crystals. Direct measurements of Zak's phase were
performed for cold atoms in optical lattices~\cite{Atala_Nature-Physics_2013}.

Much of the mathematical complexity related to Wannier--Stark states is avoided
in finite systems, where the electron motion is restricted. In this case, it is
common to refer to the exact length-gauge eigenstates of $\hat{H}_\mathrm{LG}$
as Wannier--Stark states (without neglecting interband transitions by evaluating
Wannier--Bloch states). These states have properties similar to those of the
states introduced by Wannier. One of the most important differences is that the
exact eigenstates of a Hamiltonian in one spatial dimension may not be
degenerate. While the states defined by \eqref{eq:WS_energies} are strictly
linear functions of $F_\mathrm{L}$, and, for a certain value of $F_\mathrm{L}$,
some of these energies $\En_{i_1,\ell_1}^\mathrm{WS}$ and
$\En_{i_2,\ell_2}^\mathrm{WS}$ may be equal to each other, the corresponding
exact energies of a confined quantum system will have avoided crossings
(anticrossings). An example of such a Wannier--Stark ladder is shown in
Fig.~\ref{fig:WS_ladder}.

\subsection{Accelerated Bloch states}
\label{sec:Houston}

In section \ref{sec:Wannier}, we saw that the instantaneous eigenstates of the
length-gauge Hamiltonian serve as a convenient basis for developing approximate
solutions to the TDSE. In situations where the difficulties related to the
length gauge outweigh its advantages, the velocity gauge may be a better choice
for either numerical or analytical approximations, and the instantaneous
eigenstates of the velocity-gauge Hamiltonian \eqref{eq:H_VG} may provide a more
useful time-dependent basis.

Let $\phi_{i,\mathbf{k}}$ be a Bloch state with a band index $i$ and a
crystal momentum $\mathbf{k}$:
\begin{equation}
  \label{eq:Bloch_eigenstate}
  \left(\frac{\hat{\mathbf{p}}^2}{2 m} + U(\mathbf{r})\right) \phi_{i,\mathbf{k}} =
  \En_i(\mathbf{k}) \phi_{i,\mathbf{k}}.
\end{equation}
In the coordinate representation,
$\phi_{i,\mathbf{k}}(\mathbf{r})$ is
a product of a plane wave and a lattice-periodic envelope function:
\begin{equation}
  \label{eq:Bloch_decomposition}
  \phi_{i,\mathbf{k}}(\mathbf{r}) = \ee^{\iu \mathbf{k} \mathbf{r}} u_{i,\mathbf{k}}(\mathbf{r}),
\end{equation}
where $u_{i,\mathbf{k}}\left(\mathbf{r} + \mathbf{R}\right) =
u_{i,\mathbf{k}}(\mathbf{r})$ for all $\mathbf{R}$ from the Bravais lattice. Let
us now consider the instantaneous eigenstates of $\hat{H}_\mathrm{VG}$ in the
presence of a homogeneous external field:
\begin{equation}
  \label{eq:Houston_equation}
  \left(\frac{\left[\hat{\mathbf{p}} + e \mathbf{A}_\mathrm{L}(t)\right]^2}{2 m} + U(\mathbf{r})\right) \varphi(t) =
  \widetilde{\En}(t) \varphi(t).
\end{equation}
Since the Hamiltonian is periodic in space, the Bloch theorem is applicable.
Equation~\eqref{eq:Houston_equation} has the same form as
\eqref{eq:Bloch_eigenstate}, the momentum operator being substituted with
$\hat{\mathbf{p}} + e \mathbf{A}_\mathrm{L}(t)$. The requirement that the
solutions of \eqref{eq:Houston_equation} satisfy the Born--von K\'{a}rm\'{a}n
boundary conditions yields~\cite{Krieger_1986_33_5494}
\begin{align}
  \label{eq:Houston}
  \varphi_{i,\mathbf{k}_0}(\mathbf{r},t) &= \exp\left[-\frac{\iu}{\hbar} e \mathbf{A}_\mathrm{L}(t) \mathbf{r}\right] \phi_{i,\mathbf{k}(t)}(\mathbf{r}),\\
  \widetilde{\En}_{i,\mathbf{k}_0}(t) &= \En_i\bigl(\mathbf{k}(t)\bigr),
\end{align}
where the time-dependent crystal momentum
\begin{equation}
  \label{eq:time_dependent_k}
  \mathbf{k}(t) = \mathbf{k}_0 + \frac{e}{\hbar} \mathbf{A}_\mathrm{L}(t)
\end{equation}
satisfies the acceleration theorem: $\hbar\,\de \mathbf{k}/\de t = -e
\mathbf{F}_\mathrm{L}(t)$. Here, $\mathbf{k}_0$ is the initial crystal momentum,
which the electron possessed prior to the interaction with the laser pulse.

The states $\varphi_{i,\mathbf{k}_0}(\mathbf{r},t)$ are called accelerated Bloch
states or Houston functions. We note that there is a close analogy between these
functions and Volkov solutions of the TDSE~\cite{Volkov_1935_ZP_94_250}, which
are frequently used for describing the strong-field ionization of atoms and
molecules~\cite{Jones_1977_PRB_16_2466}.

To use Houston functions as a basis for solving the TDSE, it is convenient to
make the ansatz
\begin{equation}
  \label{eq:Houston_ansatz}
  \psi_{\mathbf{k}_0}(\mathbf{r},t) = \sum_i \alpha_{i,\mathbf{k}_0}(t)
  \exp\left[-\frac{\iu}{\hbar} \int_{t_0}^t \de t'\, \En_i\bigl(\mathbf{k}(t')\bigr)\right]
  \varphi_{i,\mathbf{k}_0}(\mathbf{r},t),
\end{equation}
which leads to the following system of differential
equations~\cite{Bychkov_1970_JETP_31_928,Krieger_1986_33_5494}:
\begin{equation}
  \label{eq:Kriegers_equations}
  \frac{\de}{\de t} \alpha_{i,\mathbf{k}_0}(t) = \frac{e}{\iu \hbar} \sum_j \alpha_{j,\mathbf{k}_0}(t)
  \mathbf{F}_\mathrm{L}(t) \cdot \Berry_{ij}\bigl(\mathbf{k}(t)\bigr)
  \exp\left[\frac{\iu}{\hbar} \int_{t_0}^t \de t' \varDelta \En_{ij}\bigl(\mathbf{k}(t')\bigr)\right].
\end{equation}
Here, $\varDelta \En_{ij}(\mathbf{k}) = \En_{i}(\mathbf{k}) - \En_{j}(\mathbf{k})$, and the matrix elements
\begin{equation}
  \label{eq:Blount}
  \Berry_{ij}(\mathbf{k})
\equiv \melement{i,\mathbf{k}}{\iu \nabla_\mathbf{k}}{j,\mathbf{k}}_\mathrm{cell}
= \frac{\iu}{\Omega} \int_\Omega \de^3 r\,
  u_{i,\mathbf{k}}^*(\mathbf{r}) \nabla_\mathbf{k} u_{j,\mathbf{k}}(\mathbf{r})
\end{equation}
describe the optical transitions between bands, where the integration is
performed over the volume $\Omega$ of a unit cell.

The relation between $\Berry_{ij}(\mathbf{k})$ and the momentum matrix elements
\begin{equation}
  \label{eq:momentum_matrix}
  \mathbf{p}_{ij}(\mathbf{k})
\equiv \melement{i,\mathbf{k}}{\hat{\mathbf{p}}}{j,\mathbf{k}}_\mathrm{cell}
= -\frac{\iu\hbar}{\Omega} \int_\Omega \de^3 r\,
  u_{i, \mathbf{k}}^*(\mathbf{r}) \nabla_\mathbf{r} u_{j, \mathbf{k}}(\mathbf{r})
\end{equation}
is given by
\begin{equation}
  \label{eq:Blount_and_momentum}
  \Berry_{i \ne j}(\mathbf{k}) = \frac{\iu \hbar \mathbf{p}_{ij}(\mathbf{k})}
  {m_0 \varDelta \En_{ij}(\mathbf{k})},
\end{equation}
where $m_0$ is the electron rest mass. This expression can only be applied
as long as $\varDelta \En_{ij}(\mathbf{k})$ in the denominator
is not equal to zero. The case of degenerate bands presents additional
mathematical
challenges~\cite{Culcer_2005_PRB_72_085110,Xiao_2010_RMP_82_1959,Foreman_2000_JPCM_12_R435};
in particular, the transition matrix elements $\Berry_{ij}(\mathbf{k})$ are singular at
degeneracies~\cite{Foreman_2000_JPCM_12_R435}.

For reference, we also give the relation between $\Berry_{ij}(\mathbf{k})$ and
the matrix elements of the position operator between the Bloch
functions~\cite{Blount_1962,Virk_2007_PRB_76_035213,Gu_PRB_2013_87_125301}:
\begin{equation}
  \label{eq:Blount_and_coordinate}
  \melement{i,\mathbf{k}'}{\hat{\mathbf{r}}}{j,\mathbf{k}}_\infty
= \int_{\mathbb{R}^3} \de^3r\, \phi_{i,\mathbf{k}'}^*(\mathbf{r}) \mathbf{r} \phi_{j,\mathbf{k}}(\mathbf{r})
= \left[\iu\delta_{ij}\nabla_\mathbf{k} + \Berry_{ij}(\mathbf{k})\right] \delta(\mathbf{k} - \mathbf{k}').
\end{equation}

In the case $\alpha_{i,\mathbf{k}_0}(t_0)=\delta_{ij}$, where an electron is
in band $j$ before the external field is turned on, a
simple approximate solution to \eqref{eq:Kriegers_equations} is~\cite{Krieger_1986_33_5494}
\begin{equation}
  \label{eq:Krieger_approximate}
  \alpha_{i \ne j,\mathbf{k}_0}(t) \approx \frac{e}{\iu\hbar} \int_{t_0}^t \de t'\,
  \mathbf{F}_\mathrm{L}(t') \cdot \Berry_{ij}\bigl(\mathbf{k}(t')\bigr)
  \exp\left[\frac{\iu}{\hbar} \int_{t_0}^{t'} \de t'' \varDelta \En_{ij}\bigl(\mathbf{k}(t'')\bigr)\right],
\end{equation}
provided that the excitation probabilities are small ($|\alpha_{i,\mathbf{k}_0}|^2 \ll 1$ for $i \ne j$).

This equation is a good starting point for numerous analytical approximations.
In the case of a constant external field ($\mathbf{F}_\mathrm{L} =
\mbox{const}$), it is convenient to rewrite the right-hand side as an integral
over the crystal momentum $\mathbf{k}$. Alternatively, the integral over time
can be approximately evaluated using the saddle-point method, which is especially useful in the
case of a monochromatic external field. Such approximations can be used to
obtain analytical expressions for the rate of strong-field-induced transitions
between valence and conduction bands of a dielectric or a semiconductor. These
transitions belong to the most important strong-field effects in solids, and
they are discussed in the next subsection.

\subsection{Nonresonant interband transitions}
\label{sec:interband}

In 1928, Zener~\cite{Zener_1928_PRSLA_145_523} used semiclassical arguments to show
that a constant external field $F$ makes
valence-band electrons of a dielectric tunnel to the conduction band at a rate
(per unit volume)
\begin{equation*}
  \Gamma_\mathrm{Zener} = \frac{e |F| a}{2 \pi
    \hbar} \exp\left[ -\frac{\pi}{2} \frac{m^{1/2} E_\mathrm{g}^{3/2}}{e
      \hbar |F|} \right].
\end{equation*}

The prefactor of the exponential function was found to be rather sensitive to a
chosen method of approximation, but all such methods yield the same argument of
the exponential function. For example, Kane derived~\cite{Kane_1959_JPCS_12_181}
the tunneling rate to be equal to
\begin{equation*}
\Gamma_\mathrm{Kane} = \frac{e^2 F^2 m^{1/2}}{18 \pi \hbar^2
  E_\mathrm{g}^{1/2}}
  \exp\left[-\frac{\pi}{2} \frac{m^{1/2} E_\mathrm{g}^{3/2}}{e
  \hbar |F|} \right],
\end{equation*}
while Keldysh obtained~\cite{Keldysh_1965_JETP_20_1307}
\begin{equation}
  \label{eq:Keldysh_rate}
  \Gamma_\mathrm{Keldysh} =
  \frac{2 |e F|^{5/2} m^{1/4}}{9 \pi^2 \hbar^{3/2} E_\mathrm{g}^{5/4}}
  \exp\left[ -\frac{\pi}{2} \frac{m^{1/2} E_\mathrm{g}^{3/2}}{e \hbar |F|} \right].
\end{equation}
It is common to state that such expressions for the tunneling rate have an
``exponential accuracy''~\cite{Popov_Physics-Uspekhi_2004_47_855}.

If the external field is not constant but oscillating at a constant frequency,
interband transitions may also occur as a result of absorbing a number of
photons sufficient to overcome the band gap. It must be emphasized that
there is no sharp distinction between
nonresonant interband tunneling and multiphoton transitions.
These are two asymptotic cases of interband excitations, which are distinguished
by the Keldysh parameter
\eqref{eq:Keldysh_gamma}: $\gamma_\mathrm{K} \gg 1$ for multiphoton excitations
and $\gamma_\mathrm{K} \ll 1$ for tunneling. Equation~\eqref{eq:Keldysh_rate} is
valid in the latter case, where the laser field is strong and its frequency is
small. In the intermediate regime ($\gamma_\mathrm{K} \sim 1$), it is impossible
to unambiguously distinguish between contributions from multiphoton absorption
and tunneling. This is reminiscent of the situation in atomic physics
where the electron motion under the potential barrier (tunneling) is known to be
important even in the multiphoton regime~\cite{Ivanov_JMO_2005_52_165}.
Keldysh also derived a more general expression for the transition rate averaged
over a laser cycle, which is applicable in all the three
regimes~\cite{Keldysh_1965_JETP_20_1307}:
\begin{equation}\label{eq:GKF}
  \Gamma_\mathrm{GKF} = \frac{2\omega_\mathrm{L}}{9\pi}
  \left[\frac{m \omega_\mathrm{L}}{\hbar \beta}\right]^{3/2}
  Q( \gamma_\mathrm{K}, \widetilde{N})
  \exp\left[
   -\pi\lfloor \widetilde{N} + 1 \rfloor
		\frac{K(\beta) - E(\beta)}{E(\alpha)}
	\right],
\end{equation}
\begin{multline*}
Q(\gamma_\mathrm{K}, \widetilde{N}) =
	\left[\frac{\pi}{2 K(\alpha)}\right]^{1/2}
	\sum_{n = 0}^{\infty} \exp\left[
		- \frac{\pi n [K(\beta) - E(\beta)]}{E(\alpha)}
	\right]
\\
\times
	\Phi\left\{\left[\frac{\pi^2 (2 \lfloor \widetilde{N} + 1 \rfloor - 2 \widetilde{N} + N) }{
		2 K(\alpha) E(\alpha)
	}\right]^{1/2}\right\},
\end{multline*}
\begin{equation*}
	\alpha = (1+\gamma_\mathrm{K}^2)^{-1/2}
,\quad
	\beta = \gamma_\mathrm{K} \alpha
,\quad
	\widetilde{N} = \frac{\widetilde{E}_\mathrm{g}}{\hbar\omega_\mathrm{L}}
,\quad
	\widetilde{E}_\mathrm{g} = \frac{2 E(\alpha)}{\pi \beta} E_\mathrm{g}
,\quad
  N = \frac{E_\mathrm{g}}{\hbar\omega_\mathrm{L}}
.
\end{equation*}
Here, $\widetilde{E}_\mathrm{g}$ is an effective ionization potential, the
functions $K(z)$ and $E(z)$ are the complete elliptic integrals of first and
second kind, $\Phi(z)$ is the Dawson function, and $\lfloor x \rfloor$ denotes
the integer part of $x$. We refer to \eqref{eq:GKF} as the general Keldysh
formula (GKF).

\begin{figure}[t]
  \centering
  \includegraphics{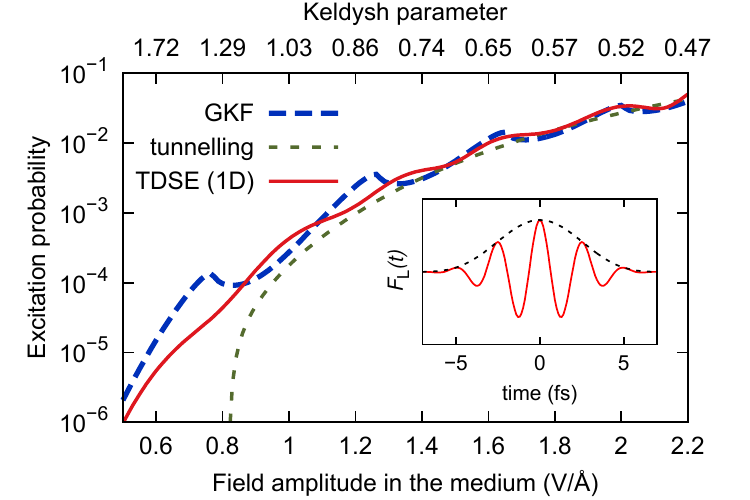}
  \caption{%
    The probability that a 4-fs laser pulse excites a valence-band electron of
    SiO$_2$ to one of the conduction bands. The laser pulse, shown in the inset,
    has a central wavelength of $\lambda_\mathrm{L}=800$~nm. The dashed
    curves show probabilities evaluated by integrating analytical excitation
    rates: $1 - \exp[-\int \Gamma(t)\,\de t]$. For the curve marked as 'GKF',
    the general Keldysh formula \eqref{eq:GKF} was used as a cycle-averaged
    excitation rate; evaluating $\Gamma_\mathrm{GKF}(t)$ we substituted
    $F_\mathrm{L}$ with the envelope of the laser pulse (dashed line in the
    inset). For the curve marked as 'tunneling', we used the quasistatic
    approximation in the tunneling limit $\Gamma(t) =
    \Gamma_\mathrm{Keldysh}(t)$ by substituting $F_\mathrm{L}$ with the electric
    field $F_\mathrm{L}(t)$ in \eqref{eq:Keldysh_rate}. The solid curve is a
    numerical result obtained by solving the TDSE in a 1D
    model~\cite{Kruchinin_2013_PRB_87_115201}. The transition rates
    \eqref{eq:Keldysh_rate} and \eqref{eq:GKF} were multiplied with constant
    factors to match the TDSE result for $\gamma_\mathrm{K} \lesssim 1$.}
  \label{fig:Keldysh}
\end{figure}

Are these formulas, which were obtained decades ago for a monochromatic external
field, still useful in the case of few-cycle laser pulses? To address this
question, we present, in Fig.~\ref{fig:Keldysh}, the outcomes of a simulation
where a 4-fs 800-nm laser pulse interacts with a one-dimensional model medium
that has properties resembling those of
$\alpha$-quartz~\cite{Kruchinin_2013_PRB_87_115201}: a band gap of $E_\mathrm{g} = 9$~eV,
a lattice period of 5~\AA{}, and a reduced mass of $m = 0.38 m_0$.
To compare the numerically evaluated excitation probabilities (solid
curve) with the GKF predictions, we evaluated the excitation rate
$\Gamma_\mathrm{GKF}(t)$ using the real-valued pulse envelope in the place of
the electric field $F_\mathrm{L}$, which enters \eqref{eq:GKF} via $\gamma_\mathrm{K}$.
From $\Gamma_\mathrm{GKF}(t)$, we estimated the excitation probability as $p_\mathrm{GKF}
\propto 1 - \exp[-\int \Gamma_\mathrm{GKF}(t)\,\de t]$ (thick dashed curve). The
overall agreement is surprisingly good, given the fact that the laser pulse is
shorter than two optical periods. Both the numerical and GKF results exhibit an
oscillatory behavior, which appears due to closing and opening of multiphoton channels.
It is analogous to the channel closing phenomenon in atomic
physics~\cite{Story_1994_PRA_49_3875,Paulus_2001_PRA_64_021401,Kopold_2002_JPB_35_217}.
The thin dashed curve in Fig.~\ref{fig:Keldysh} represents the excitation
probability evaluated with \eqref{eq:Keldysh_rate}. Since
$\Gamma_\mathrm{Keldysh}$ is the tunneling rate for a constant field, we used
$F_\mathrm{L}(t)$ (rather than the pulse envelope) to evaluate the excitation
probability by the laser pulse. This procedure is known as the quasistatic
approximation~\cite{Ivanov_CP_2013_414_3}. In this example, the tunneling
formula \eqref{eq:Keldysh_rate} is inaccurate for $F_\mathrm{L} \lesssim 1.2$~V/\AA{}, which corresponds to $\gamma_\mathrm{K} \gtrsim 0.9$.

The pioneering work by Keldysh was followed by numerous investigations. A few
that we would like to point out here are analytical results obtained using the
adiabatic approach, where parity selection rules were accounted
for~\cite{Bychkov_1970_JETP_31_928,Minasian_1986_PRB_34_963}, derivation of expressions for
arbitrary $N$-photon transition probabilities~\cite{Kovarskii_1971_PSSB_45_47},
and development of the Keldysh-like theory for cosine-shaped
bands~\cite{Gruzdev_2007_PRB_75_205106}.

\section{Strong-field-driven electron dynamics in crystals}
\label{sec:main_phenomena}

\subsection{A numerical example}

A strong electric field drives interband transitions, it accelerates charge
carriers, and it can also cause transient changes in the optical properties of a
medium without necessarily exciting electrons to conduction bands. Combined with
controlled optical fields, these three classes of physical phenomena enable the
manipulation of the electric and optical properties of a medium over time
intervals much shorter than a period of optical oscillations. The main examples
of such controlled fields are laser pulses with the stabilized carrier--envelope
phase (CEP) and optical waveforms~\cite{Wirth_Science_2011}.

\begin{figure}[t]
  \centering
  \includegraphics{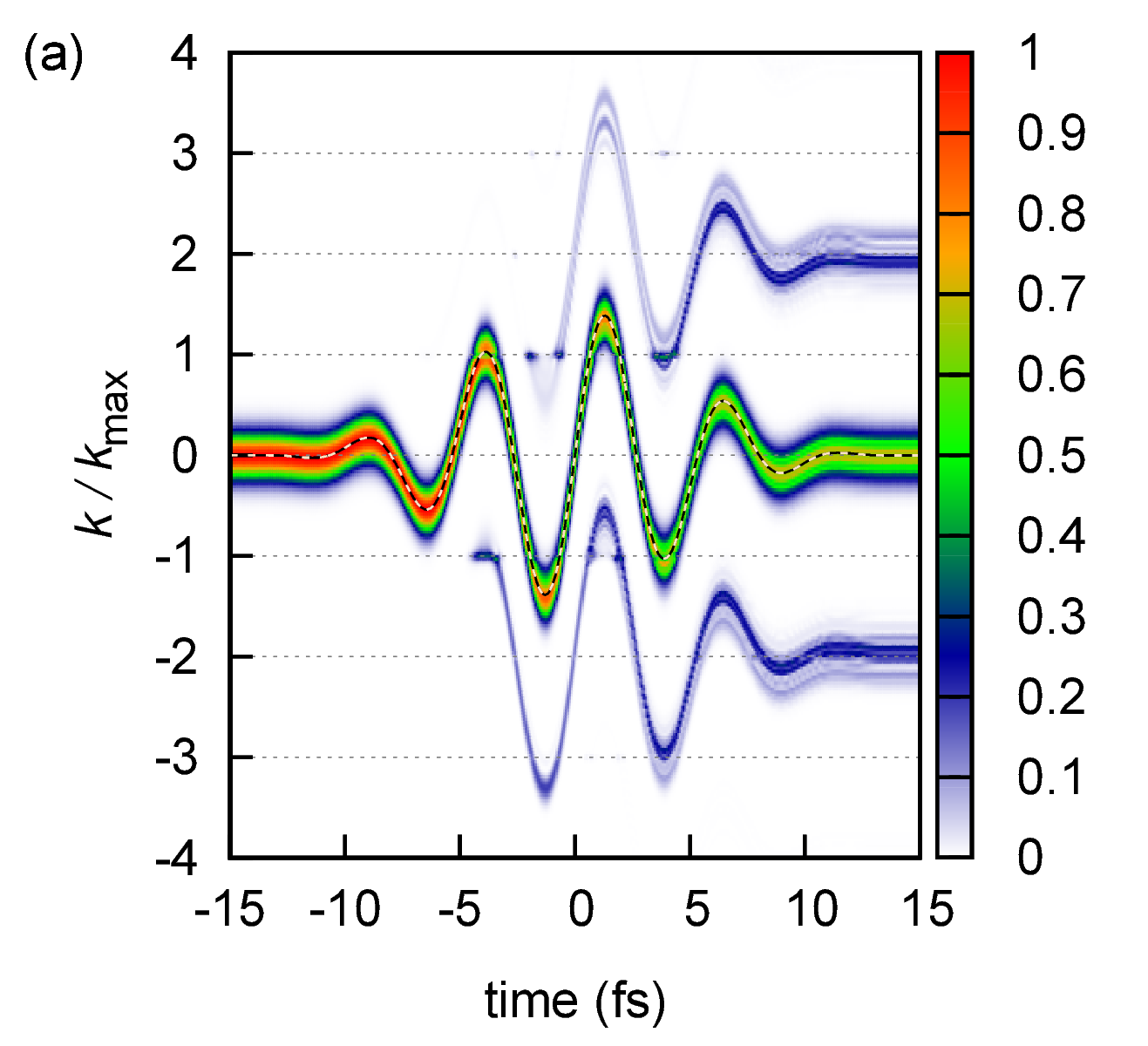}
  \includegraphics{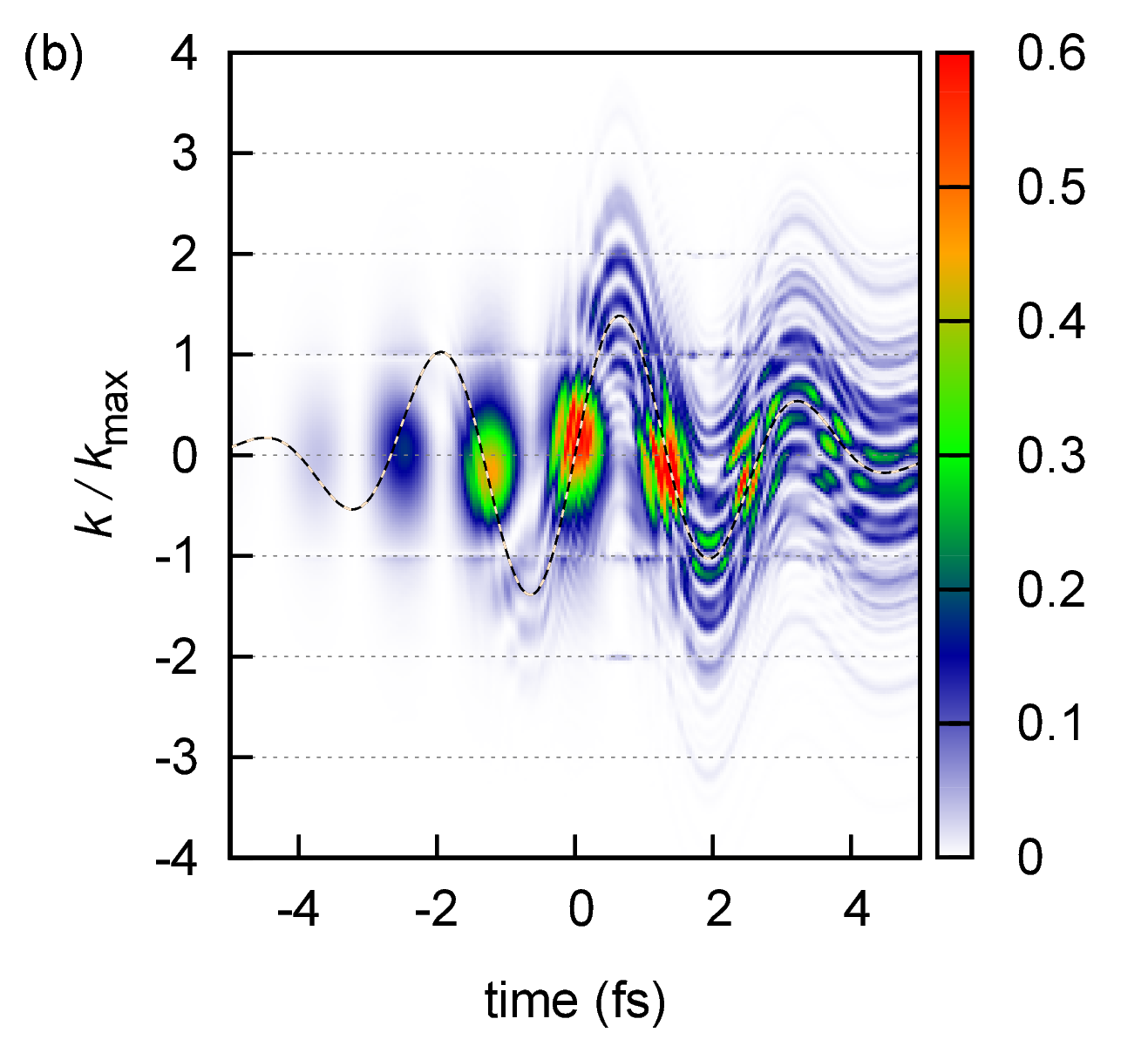}
  \caption{The time-dependent population distribution of conduction-band electrons in the
    presence of a few-cycle laser pulse. These plots were obtained by solving
    \eqref{eq:Kriegers_equations} for a 1D model of SiO${}_2$ with the same lattice
    potential as that used in~\cite{Kruchinin_2013_PRB_87_115201}. The normalized
    conduction-band population ($\max n(t,k) = 1$) is plotted against time and
    crystal momentum $k$ in the extended-zone scheme, where the range of crystal
    momenta covered by the $n$-th lowest conduction band is $n-1 \le |k| /
    k_\mathrm{max} \le n$ with $k_\mathrm{max} = \pi/a$. The dashed curves show
    $A_\mathrm{L} / k_\mathrm{max}$. (a) An electron wave packet is initially
    placed in the lowest conduction band $\left(\alpha_{n_0,k}(t_0) = \exp[-10
    (k/k_\mathrm{max})^2]\right)$; an 8-fs laser pulse with a central wavelength of
    $\lambda_\mathrm{L} = 1.6$~\textmu{}m and a peak field of $F_\mathrm{L} =
    0.7$~V/\AA{} is sufficiently strong to accelerate the electrons
    out of the first Brillouin zone. (b) All electrons are initially in the
    valence bands; a 4-fs pulse with $\lambda_\mathrm{L} = 800$~nm and
    $F_\mathrm{L} = 1.4$~V/\AA{} excites them to the conduction
    bands, where they are accelerated by the laser field.}
  \label{fig:tk_analysis}
\end{figure}

We begin this section by illustrating such effects in a simulation where the
TDSE was solved in one spatial dimension for a periodic potential. In
Fig.~\ref{fig:tk_analysis}, we show $|\alpha_{i,k(t)}(t)|^2$ obtained by solving
the Houston-basis equations~\eqref{eq:Kriegers_equations} and representing the
conduction-band populations in the extended-zone scheme, where the range of
crystal momenta covered by the $n$-th conduction band ($n \ge 1$) is $n-1 \le
|k| / k_\mathrm{max} \le n$ with $k_\mathrm{max} = \pi/a$. According to the
acceleration theorem, the ballistic motion of an electron wave packet is
described by $\hbar \de \langle k \rangle/\de t = -e F_\mathrm{L}(t)$, where
$\langle k \rangle$ is the mean crystal momentum. A continuous change of
$\langle k \rangle$ at the borders $|k| / k_\mathrm{max} = n$ corresponds to
transitions between different conduction bands. Such interband transitions are
closely related to Landau--Zener transitions, as they occur at crystal momenta
where the energy gap between two adjacent bands is particularly small. If a
charge carrier remains within its current band as it crosses a Brillouin zone
edge, its crystal momentum changes abruptly. This can be interpreted as a
Bragg-like reflection of an electron wave off the crystal lattice.
Fig.~\ref{fig:tk_analysis} shows that both Bragg reflections and interband
transitions play an important role when the electron motion is driven by a
near-infrared field.

In Fig.~\ref{fig:tk_analysis}(a), the electron wave packet is initially placed
in the lowest conduction band of the 1D model of a solid, and even though the laser pulse
with $\lambda_\mathrm{L}=1.6$~\textmu{}m is strong enough to accelerate
electrons out of the first Brillouin zone, its amplitude is insufficient
to induce transitions between conduction and valence bands. The figure illustrates the
importance of transitions between different conduction bands, and shows that
they predominantly occur at $k=\pi n / a$, $n \in \mathbb{Z}$.

For the simulations presented in Fig.~\ref{fig:tk_analysis}(b), all electrons
were initially placed in the valence bands. The laser pulse had the same peak
value of the vector potential as in Fig.~\ref{fig:tk_analysis}(a), but it had a
shorter wavelength $\lambda_\mathrm{L}=800$~nm, so that the electric
field of the pulse was twice as strong, and it was strong enough to excite
electrons from the uppermost valence band. One can see that the population of
conduction bands does not constantly increase with time, as one would expect
from rate models discussed in section~\ref{sec:interband}. Instead, there is a
transient increase of the population at the extrema of the electric field (at
the zero crossings of the vector potential). Such excitations are called
``virtual''~\cite{Yablonovitch_1989_PRL_63_976} because they only exist as long
as the external field is present.
If a laser pulse is so weak that the final excitation
probability can be neglected, such virtual excitations represent the distortion
of bound states of the crystal---the distorted valence-band states have nonzero
projections onto the field-free conduction-band Bloch functions. In the
strong-field regime, no clear distinction between the ``virtual'' and ``real''
excitation can be made. Nevertheless, in both weak- and strong-field regimes,
the transient increase of conduction-band population has an effect on
experimentally observable quantities~\cite{Yablonovitch_1989_PRL_63_976,%
Kuznetsov_1993_PRB_48_10828}.

Currently, there is no measurement technique that would reveal all the details
of strong-field-driven electron dynamics like those illustrated by
Fig.~\ref{fig:tk_analysis}. However, time-resolved measurement techniques do
provide indirect access to this information. Some evidence of subcycle dynamics
in the strong-field excitation of electron in SiO$_2$ was presented
in~\cite{Gertsvolf_2010_JPB_43_131002} by measuring the polarization rotation of
an elliptically polarized pulse transmitted through a thin glass plate, as well
as in~\cite{Mitrofanov_2011_PRL_106_147401} using a noncollinear pump--probe
measurement scheme. Bragg-like scattering of electrons was found to contribute
to the generation of nonperturbative high-order harmonics in solid
samples~\cite{Ghimire_2011_NP_7_138,Schubert_2014_NP_8_119}. The reversible
field-induced change of absorption in the extreme ultraviolet spectral range was
observed by probing the effect of an intense near-infrared field on a thin
silica plate using an attosecond pulse of extreme ultraviolet radiation as a
probe~\cite{Schultze_2013_Nature_493_75}. A
subcycle turn-on of electric current in a dielectric and its manipulation with
CEP-stabilized pulses was demonstrated by measuring the residual polarization
induced by the laser light~\cite{Schiffrin_2013_Nature_493_70}. This last effect
is most closely related to the topic of this chapter, so it is discussed in more
detail in the following section.

\subsection{Ultrafast injection and control of current in dielectrics}
\begin{figure}[t]
  \centering
  \includegraphics[width=55mm]{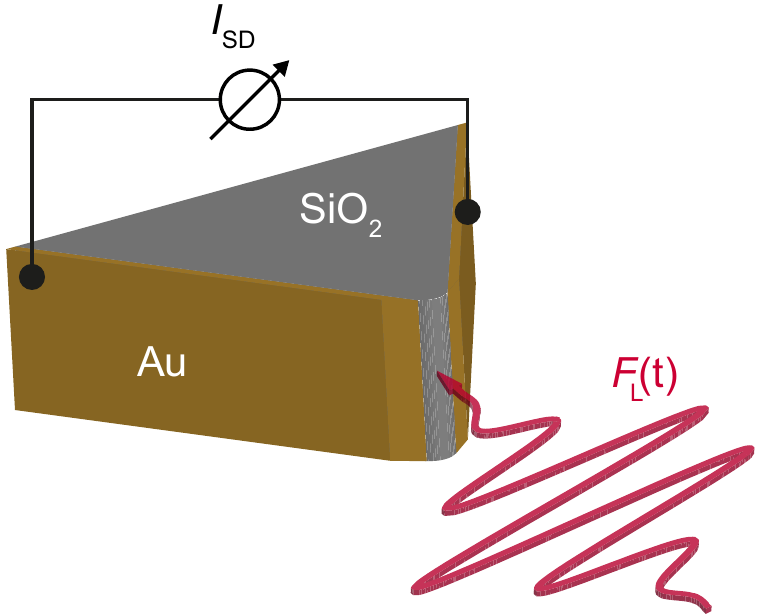}
  \caption{A schematic representation of the experimental arrangement used
    in~\cite{Schiffrin_2013_Nature_493_70}. An intense few-cycle laser pulse
    induces electric current in the dielectric (SiO${}_2$) placed between two
    gold electrodes. The measured signal is proportional to the net charge
    displaced by the laser pulse, and it is controlled by the carrier--envelope
    phase of the laser pulse.}
    \label{fig:exp_setup}
\end{figure}

According to Fig.~\ref{fig:tk_analysis}(b), an optical field that is strong
enough to excite valence-band electrons of a dielectric to its conduction bands
is also strong enough to significantly accelerate the created charge carriers,
thus driving electric current. One of the most important findings
in~\cite{Schiffrin_2013_Nature_493_70} was that such electric current can be
turned on within a fraction of a half-cycle of a short intense laser pulse. This
was demonstrated by irradiating a SiO$_2$ sample placed between two gold electrodes
with CEP-stabilized laser pulses (see Fig.~\ref{fig:exp_setup}) and measuring
the current induced by the pulses in an external circuit. It was found that a
short laser pulse with the electric field directed perpendicularly to the
electrodes was able to leave the sample in a polarized state, implying that a
certain electric charge was displaced by the pulse. By varying the CEP of the
pulse, it was possible to control the amount of the displaced charge. This fact
alone already suggests that the observed effect should be controlled by the
electric field (rather than the envelope) of the laser pulse, but the most
convincing evidence for the subcycle turn-on of electric current was provided by
pump--probe measurements, where the sample was irradiated by a pair of pulses:
an intense ``injection pulse'' polarized parallel to the electrodes and a
relatively weak ``drive'' pulse polarized perpendicularly to them. By observing
how the displaced charge depends on the delay between the two pulses, it was
possible to conclude that the injection pulse makes the SiO$_2$ sample
conductive within a time interval $\lesssim 1$~fs. These measurements were well
reproduced by simulations: the one-dimensional tight-binding simulations in the
original paper~\cite{Schiffrin_2013_Nature_493_70}, a model that used a
one-dimensional pseudopotential~\cite{Kruchinin_2013_PRB_87_115201}, and recent
\textit{ab initio} three-dimensional simulations~\cite{Wachter_PRL_2014}.
Nevertheless, these observations permit several interpretations, which we
present in the rest of this section.

\subsubsection*{Semiclassical interpretation}
The most intuitive interpretation is based on the observation that the
strong-field-driven motion of an electron wave packet in the conduction bands of
a dielectric largely obeys the acceleration theorem (see
Fig.~\ref{fig:tk_analysis}(a))---apart from interband transitions and occasional
Bragg scattering, which splits and reunites electron wave packets, each
wave packet moves as a classical particle with an effective mass that depends on
the mean crystal momentum. This suggests that the residual polarization induced
by the laser pulse may be interpreted in semiclassical terms. A rigorous
approach to this interpretation consists in writing the current density averaged
over a unit cell
\begin{equation}
  \label{eq:current}
  \mathbf{J}(t) = \int_\mathrm{BZ} \frac{\de^3 k_0}{(2 \pi)^3} \mathbf{j}_{\mathbf{k}_0}(t)
\end{equation}
in the basis of Houston functions, where the contribution from an electron with
an initial crystal momentum $\mathbf{k}_0$ is given by
\begin{multline}
  \label{eq:Houston_current}
  \mathbf{j}_{\mathbf{k}_0}(t) = -e \sum_i \left| \alpha_{i,\mathbf{k}_0}(t) \right|^2 
    \mathbf{v}_i\bigl(\mathbf{k}(t)\bigr)\\
    -\frac{2 e}{m_0} \sum_{i,j<i} \mbox{Re}\left\{
      \alpha_{i,\mathbf{k}_0}^*(t) \alpha_{j,\mathbf{k}_0}(t)
      \mathbf{p}_{ij} \bigl(\mathbf{k}(t)\bigr)
      \exp\left[\frac{\iu}{\hbar} \int_{t_0}^t \de t' \varDelta \En_{ij}\bigl(\mathbf{k}(t')\bigr)\right]
    \right\}.
\end{multline}
Here, $\mathbf{k}(t)$ is defined by \eqref{eq:time_dependent_k}, $m_0$ is the
electron rest mass, $\mathbf{p}_{ij}(\mathbf{k})$ are the momentum matrix
elements \eqref{eq:momentum_matrix}, and $\mathbf{v}_i(\mathbf{k}) =
\nabla_\mathbf{k} \En_i(\mathbf{k})/\hbar = \mathbf{p}_{ii}(\mathbf{k})/m_0$ is the
group velocity in band $i$. The first sum on the right-hand side of
\eqref{eq:Houston_current} is responsible for the current due to the ballistic
motion of charge carriers. In the semiclassical interpretation, the contribution
from this term to the residual polarization (displaced charge density)
\[ \mathbf{P}(t_\mathrm{max}) = \int_{-\infty}^{t_\mathrm{max}} \mathbf{J}(t)\,\de
t \] is assumed to be much larger than that from the second sum, which describes
effects related to interband coherences.

In the tunneling regime ($\gamma_\mathrm{K} \lesssim 1$), charge carriers are
predominantly created at the extrema of the electric field, each of which
launches an electron wave packet. For a wave packet launched at a
time $t_0$, it is convenient to introduce a \emph{semiclassical displacement}:
\begin{equation}
  \label{eq:displacement}
  \mathbf{s}(t_0) = \int_{t_0}^{t_\mathrm{max}} \mathbf{v}\bigl(\mathbf{k}(t)\bigr)\,\de t,
\end{equation}
where the group velocity $\mathbf{v}(\mathbf{k})$ should correspond to the most
probable quantum path of the wave packet in reciprocal space. As long as
$\mathbf{k}(t)$ is not limited to the first Brillouin zone, this approach is
most useful if the probabilities of Bragg scattering at the edges of the
Brillouin zone are either negligibly
small 
or close to 100\%. 
The contribution from each wave packet to the final polarization
$\mathbf{P}(t_\mathrm{max})$ is the product of the charge carried by the wave
packet and its semiclassical displacement $\mathbf{s}(t_0)$.

This kind of semiclassical analysis explains some outcomes of numerical
simulations~\cite{Foldi_2013_NJP_15_063019}. In particular, it explains the
observation that, for moderate laser intensities, the residual polarization
scales as $P \propto F_\mathrm{L}^{2 N +
  1}$~\cite{Kruchinin_2013_PRB_87_115201}, where $N = E_\mathrm{g}/(\hbar
\omega_\mathrm{L})$ is the ratio of the band gap to the photon energy.
As long as the Keldysh parameter is sufficiently large to
view interband excitations as the result of absorbing $N$ photons, the
excitation probability scales as $p \propto I_\mathrm{L}^N \propto
F_\mathrm{L}^{2 N}$, $I_\mathrm{L}$ being the peak laser intensity, while the
semiclassical displacement is proportional to $F_\mathrm{L}$. Thus, the product of
the charge and the displacement is proportional to $F_\mathrm{L}^{2 N + 1}$.

\subsubsection*{Interference of multiphoton pathways}
For moderate laser intensities, where perturbation theory is expected to yield
at least qualitatively correct predictions, it is also possible to interpret the
optically controlled electric current in terms of interference between different
multiphoton excitation pathways. This interpretation is a generalization of
ideas developed in the field of coherent control, where irradiating a
semiconductor by two monochromatic laser beams with frequencies $\omega_1$ and
$\omega_2 = 2 \omega_1$ was found to induce an electric current sensitive to the
relative phase between the two beams~\cite{Fortier_2004_PRL_92_147403}. This
phase sensitivity is due to the interference between single- and two-photon
absorption processes. The density $n(\mathbf{k})$ of electrons excited to the
conduction band at a crystal momentum $\mathbf{k}$ is determined by the phase
parameter $\varDelta\varphi = 2\varphi_{\omega_1} - \varphi_{\omega_2}$.
Furthermore, $n(\mathbf{k})$ is, in general, an asymmetric function of
$\mathbf{k}$ because the transition amplitudes for the one- and two-photon
channels have different symmetries with respect to the transformation
$\mathbf{k} \to -\mathbf{k}$~\cite{Fortier_2004_PRL_92_147403}.
Injecting currents through interfering photoexcitation pathways was
investigated in theory and experiments for semiconductors
\cite{Kurizki_1989_PRB_39_3435,Atanasov_1996_PRL_76_1703,Hache_1997_PRL_78_306,
  Fraser_1999_PB_272_1999,Fortier_2004_PRL_92_147403,Zhao_2008_JAP_103_053510,
  Costa_2007_NP_3_632,Rioux_2012_PE_45_1} and molecular wires
\cite{Franco_2007_PRL_99_126802}.

\begin{figure}[!ht]
	\centering
	\includegraphics[width=0.5\textwidth]{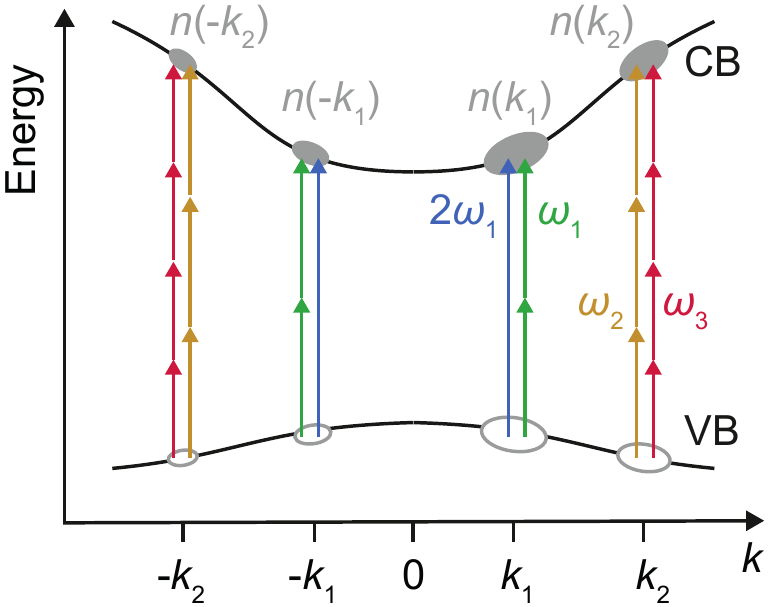}
    \caption[Current Injection via Interference of Multiphoton Excitation Channels]
    {Current injection via interference of multiphoton excitation channels.
      Charge carriers can be injected from a valence band (VB) into a conduction
      band (CB) via different multiphoton excitation pathways if the spectral
      bandwidth of the applied laser light is sufficiently large. For example,
      the excitation at crystal momenta $k_2$ and $-k_2$ may be the outcome of
      absorbing either three high-energy or four low-energy photons from the
      same ultrashort laser pulse. Quantum interference between the odd- and
      even-numbered contributions in each pathway determines the population in
      the final state. An asymmetry in the conduction-band population
      $\bigl(n(k) \ne n(-k)\bigr)$ results in a net current density is formed
      inside the material.}
	\label{fig:BASICS-coherent-control}%
\end{figure}

In the case where charge carriers are excited by a few-cycle pulse with a
central frequency that is much smaller than the band gap, the interfering
photoexcitation pathways are multiphoton excitation channels. It is convenient
to analyze them in the velocity gauge using the basis of Bloch states, where an
external homogeneous electric field only induces transitions between states with
the same the crystal momentum. Due to the large bandwidth of ultrashort laser
pulses, it is possible to make the same transition by absorbing different
numbers of photons, as it is schematically shown in
Fig.~\ref{fig:BASICS-coherent-control}.

Within this picture, the scaling law $Q \propto F_\mathrm{L}^{2 N + 1}$ may be
interpreted as a result of interference between quantum pathways that involve
absorbing $N$ and $N+1$ photons. The corresponding probability amplitudes are
proportional to $F_\mathrm{L}^{N}$ and $F_\mathrm{L}^{N+1}$, respectively. When
the two pathways interfere, the excitation probability, which is the squared
modulus of the sum of the probability amplitudes, is described by an
expression that contains the product of the two amplitudes. This product, which
is proportional to $F_\mathrm{L}^{2 N + 1}$, determines the induced electric
current. Even though these arguments by no means form a rigorous proof, and
these considerations are only applicable in the multiphoton regime, the
interpretation in terms of interfering multiphoton channels was shown to be a
plausible one, explaining not only the scaling with intensity, but also the fact
that a CEP-sensitive displaced charge can only be observed if the laser pulse is
sufficiently broadband~\cite{Kruchinin_2013_PRB_87_115201}.

\subsubsection*{Adiabatic metallization}

Another interpretation of the optical-field-induced current in dielectrics
accompanied the first publication of these experimental
results~\cite{Schiffrin_2013_Nature_493_70}. This mechanism relies on the effect
of ``adiabatic metallization'', predicted theoretically for dielectric
nanofilms~\cite{Durach_2010_PRL_105_086803,Durach_2011_PRL_107_086602}. It was
found that a strong electric field can significantly and reversibly change the
optical and electric properties of a sufficiently thin dielectric nanofilm.
During the interaction with the field, the real part of the dielectric constant
may even become negative, which is a property attributed to metals. These
effects are best understood in the basis of instantaneous eigenstates of the
length-gauge Hamiltonian, which are the Wannier--Stark states discussed at the
end of section~\ref{sec:Wannier}. Even though these results were obtained for
nanofilms, similar effects may be expected in bulk
solids~\cite{Apalkov_PRB_2012_86_165118}, provided that the localization length
of Wannier--Stark states does not exceed a few lattice sites, which is indeed
the case for field strengths $F_\mathrm{L} \gtrsim 1$~V/\AA.

\begin{figure}[t]
  \centering
  \includegraphics[width=60mm]{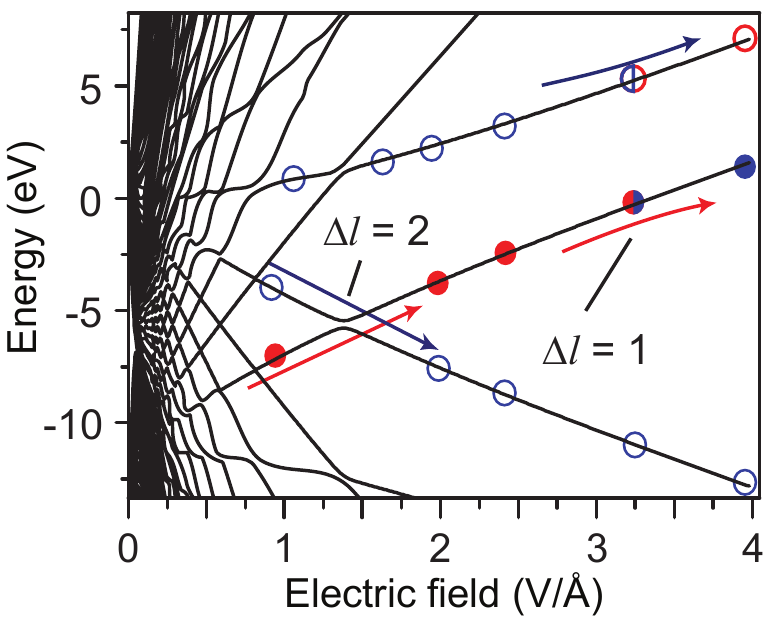}
  \caption{Eigenstates of the Wannier--Stark Hamiltonian \eqref{eq:H_LG} for a
    50-nm nanofilm of SiO${}_2$ in dependence of the electric field. The labels
    $\Delta l = 1$ and $\Delta l = 2$ indicate the anticrossings that correspond
    to Wannier--Stark states being localized one and two lattice sites apart,
    respectively. Closed circles represent occupied states, while open circles
    represent unoccupied states. The figure is adapted from
    Ref.~\cite{Schiffrin_2014_Nature}.}
  \label{fig:WS_ladder}
\end{figure}

Using the Wannier--Stark states as a time-dependent basis, the interaction with
an intense pulse can be analyzed in terms of adiabatic and diabatic transitions
at avoided crossings. This
analysis~\cite{Schiffrin_2013_Nature_493_70,Schiffrin_2014_Nature} shows that
the strong field may cause a reversible quantum transition from an insulating
state to a state with an increased conductivity. This transition occurs when the
energy gaps at avoided crossings between Wannier--Stark states become
sufficiently large for electrons to adiabatically pass such an anticrossing (see
Fig.~\ref{fig:WS_ladder}). The anticrossing gap takes its largest value when the
relevant Wannier--Stark states are localized at neighboring lattice sites, which
happens when the electric field in the medium is as strong as $F_\mathrm{crit} =
E_\mathrm{g}/(e a)$. This condition is equivalent to $\hbar \omega_\mathrm{B} =
E_\mathrm{g}$. For SiO$_2$, $F_\mathrm{crit} \approx 3$~V/\AA{},
which is above the damage threshold, but even the penultimate anticrossing,
which occurs at $F_\mathrm{crit} \approx 1.5$~V/\AA{}, was found to
have a sufficiently large energy gap. This interpretation of
optical-field-controlled current as a result of adiabatic metallization was also
supported by the observation of the optical-field-induced transient reflectivity
in the strong-field regime~\cite{Schultze_2013_Nature_493_75}.

\section{Summary and outlook}

The main message of this chapter is that the control over the electric field of
light pulses can be transformed into the control over strong-field-driven
electron dynamics, which may open attractive opportunities for both basic
research and applications in ultrafast signal processing. One example of such an
application is a solid-state CEP detector~\cite{PaaschColberg_2014_NP_8_214},
and more advanced applications may emerge in the future. Attosecond science
provides powerful tools and techniques for studying strong-field dynamics in
solids, which may lead to important insights into fundamentally important
phenomena. So far, only the first few steps have been done in this direction,
and many questions remain open. Further research is necessary to verify
conjectures made in the interpretation of recent measurements, and it is yet to
be understood how electron--electron interaction, scattering, and dephasing phenomena
affect our ability to launch and manipulate electron wave packets. One of the
most important goals is to identify those extremely nonlinear effect that are
largely reversible on a few-femtosecond time scale, as only such effect may
serve as a basis for applications in signal processing. On the experimental
side, there is a large potential for exploiting various laser sources, intense
optical waveforms, and (nano) structures designed for strong-field measurements.

\section*{Acknowledgments}
We acknowledge useful discussions with and help from A.~Schiffrin and
V.~Apalkov. We thank the Munich Centre for Advanced Photonics for support.
The work of M.I.S. and V.S.Y. was supported primarily by the Chemical
Sciences, Biosciences and Geosciences Division (grant no. DEFG02-01ER15213)
and additionally by the Materials Sciences and Engineering Division (grant
no. DE-FG02-11ER46789) of the Office of the Basic Energy Sciences, Office
of Science, US Department of Energy and by a MURI grant from the US Air
Force Office of Scientific Research.

\bibliographystyle{spphys}
\bibliography{book}

\begin{thebibliography}{10}
\providecommand{\url}[1]{{#1}}
\providecommand{\urlprefix}{URL }
\expandafter\ifx\csname urlstyle\endcsname\relax
  \providecommand{\doi}[1]{DOI \discretionary{}{}{}#1}\else
  \providecommand{\doi}{DOI \discretionary{}{}{}\begingroup
  \urlstyle{rm}\Url}\fi

\bibitem{Lenzner_1998_PRL_80_4076}
M.~Lenzner, J.~Kr\"uger, S.~Sartania, Z.~Cheng, C.~Spielmann, G.~Mourou,
  W.~Kautek, F.~Krausz, Phys. Rev. Lett. \textbf{80}, 4076 (1998)

\bibitem{Mao_APA_2004_79_1695}
S.S. Mao, F.~Quéré, S.~Guizard, X.~Mao, R.E. Russo, G.~Petite, P.~Martin,
  Appl. Phys. A \textbf{79}(7), 1695 (2004)

\bibitem{Krausz_2009_RMP_81_163}
F.~Krausz, M.~Ivanov, Rev. Mod. Phys. \textbf{81}, 163 (2009)

\bibitem{Garmire_OE_2013_21_30532}
E.~Garmire, Opt. Express \textbf{21}(25), 30532 (2013)

\bibitem{Krausz_NaturePhotonics_2014}
F.~Krausz, M.I. Stockman, Nature Photonics \textbf{8}(3), 205 (2014)

\bibitem{Ghimire_JPB_2014}
S.~Ghimire, G.~Ndabashimiye, A.D. DiChiara, E.~Sistrunk, M.I. Stockman,
  P.~Agostini, L.F. DiMauro, D.A. Reis, J. Phys. B \textbf{47}(20), 204030
  (2014)

\bibitem{Ghimire_2011_NP_7_138}
S.~Ghimire, A.D. DiChiara, E.~Sistrunk, P.~Agostini, L.F. DiMauro, D.A. Reis,
  Nature Physics \textbf{7}(2), 138 (2011)

\bibitem{Schubert_2014_NP_8_119}
O.~Schubert, M.~Hohenleutner, F.~Langer, B.~Urbanek, C.~Lange, U.~Huttner,
  D.~Golde, T.~Meier, M.~Kira, S.W. Koch, R.~Huber, Nature Photonics
  \textbf{8}(2), 119 (2014)

\bibitem{Schultze_2013_Nature_493_75}
M.~Schultze, E.M. Bothschafter, A.~Sommer, S.~Holzner, W.~Schweinberger,
  M.~Fiess, M.~Hofstetter, R.~Kienberger, V.~Apalkov, V.S. Yakovlev, M.I.
  Stockman, F.~Krausz, Nature \textbf{493}(7430), 75 (2013)

\bibitem{Schiffrin_2013_Nature_493_70}
A.~Schiffrin, T.~Paasch-Colberg, N.~Karpowicz, V.~Apalkov, D.~Gerster,
  S.~M\"uhlbrandt, M.~Korbman, J.~Reichert, M.~Schultze, S.~Holzner, J.V.
  Barth, R.~Kienberger, R.~Ernstorfer, V.S. Yakovlev, M.I. Stockman, F.~Krausz,
  Nature \textbf{493}, 70 (2013)

\bibitem{Wegener_2005}
M.~Wegener, \emph{Extreme Nonlinear Optics} (Springer, 2005)

\bibitem{Aversa_1995_PRB_52_14636}
C.~Aversa, J.E. Sipe, Phys. Rev. B \textbf{52}, 14636 (1995)

\bibitem{Resta_1998_PRL_80_1800}
R.~Resta, Phys. Rev. Lett. \textbf{80}, 1800 (1998)

\bibitem{Souza_2004_PRB_69_085106}
I.~Souza, J.~\'I\~niguez, D.~Vanderbilt, Phys. Rev. B \textbf{69}, 085106
  (2004)

\bibitem{Virk_2007_PRB_76_035213}
K.S. Virk, J.E. Sipe, Phys. Rev. B \textbf{76}, 035213 (2007)

\bibitem{Springborg_2008_PRB_77_045102}
M.~Springborg, B.~Kirtman, Phys. Rev. B \textbf{77}, 045102 (2008)

\bibitem{Houston_1940_PR_57_184}
W.V. Houston, Phys. Rev. \textbf{57}, 184 (1940)

\bibitem{Wannier_1960_PR_117_432}
G.H. Wannier, Phys. Rev. \textbf{117}, 432 (1960)

\bibitem{Wannier_1962_RMP_34_645}
G.H. Wannier, Rev. Mod. Phys. \textbf{34}, 645 (1962)

\bibitem{Voisin_1988_PRL_61_1639}
P.~Voisin, J.~Bleuse, C.~Bouche, S.~Gaillard, C.~Alibert, A.~Regreny, Phys.
  Rev. Lett. \textbf{61}, 1639 (1988)

\bibitem{Zak_1968_PRL_20_1477}
J.~Zak, Phys. Rev. Lett. \textbf{20}, 1477 (1968)

\bibitem{Wannier_1969_PR_181_1364}
G.H. Wannier, Phys. Rev. \textbf{181}, 1364 (1969)

\bibitem{Avron_1977_JMP_18_918}
J.E. Avron, J.~Zak, A.~Grossmann, L.~Gunther, J. Math. Phys. \textbf{18}(5),
  918 (1977)

\bibitem{Gluck_2002_PR_366_103}
M.~Gl\"uck, A.R. Kolovsky, H.J. Korsch, Physics Reports \textbf{366}(3), 103
  (2002)

\bibitem{Resta_2000_JPCM_12_R107}
R.~Resta, J. of Phys.: Condensed Matter \textbf{12}(9), R107 (2000)

\bibitem{Sundaram_1999_PRB_59_14915}
G.~Sundaram, Q.~Niu, Phys. Rev. B \textbf{59}, 14915 (1999)

\bibitem{Xiao_2010_RMP_82_1959}
D.~Xiao, M.C. Chang, Q.~Niu, Rev. Mod. Phys. \textbf{82}, 1959 (2010)

\bibitem{Zak_1989_PRL_62_2747}
J.~Zak, Phys. Rev. Lett. \textbf{62}, 2747 (1989)

\bibitem{King-Smith_PRB_1993}
R.D. King-Smith, D.~Vanderbilt, Phys. Rev. B \textbf{47}, 1651 (1993)

\bibitem{Atala_Nature-Physics_2013}
M.~Atala, M.~Aidelsburger, J.~Barreiro, D.~Abanin, T.~Kitagawa, Nature physics
  \textbf{9}(12), 795 (2013)

\bibitem{Krieger_1986_33_5494}
J.B. Krieger, G.J. Iafrate, Phys. Rev. B \textbf{33}, 5494 (1986)

\bibitem{Volkov_1935_ZP_94_250}
D.M. Volkov, Z. Physik \textbf{94}, 250 (1935)

\bibitem{Jones_1977_PRB_16_2466}
H.D. Jones, H.R. Reiss, Phys. Rev. B \textbf{16}, 2466 (1977)

\bibitem{Bychkov_1970_JETP_31_928}
{\relax Yu}.A. Bychkov, A.M. Dykhne, Sov. Phys. JETP \textbf{58}, 1734 (1970)

\bibitem{Culcer_2005_PRB_72_085110}
D.~Culcer, Y.~Yao, Q.~Niu, Phys. Rev. B \textbf{72}, 085110 (2005)

\bibitem{Foreman_2000_JPCM_12_R435}
B.A. Foreman, J. of Phys.: Condensed Matter \textbf{12}(34), R435 (2000)

\bibitem{Blount_1962}
E.I. Blount, in \emph{Solid State Physics: Advances in Research and
  Applications}, vol.~13, ed. by F.~Seitz, D.~Turnbull (Academic Press, 1962),
  pp. 305--373

\bibitem{Gu_PRB_2013_87_125301}
B.~Gu, N.H. Kwong, R.~Binder, Phys. Rev. B \textbf{87}, 125301 (2013)

\bibitem{Zener_1928_PRSLA_145_523}
C.~Zener, Proceedings of the Royal Society of London, Series A
  \textbf{145}(855), 523 (1934)

\bibitem{Kane_1959_JPCS_12_181}
E.O. Kane, Journal of Physics and Chemistry of Solids \textbf{12}(2), 181
  (1960)

\bibitem{Keldysh_1965_JETP_20_1307}
L.V. Keldysh, Sov. Phys. JETP \textbf{20}(5), 1307 (1965)

\bibitem{Popov_Physics-Uspekhi_2004_47_855}
V.S. Popov, Physics-Uspekhi \textbf{47}(9), 855 (2004)

\bibitem{Ivanov_JMO_2005_52_165}
M.~Ivanov, M.~Spanner, O.~Smirnova, J. Mod. Opt. \textbf{52}(2-3), 165 (2005)

\bibitem{Kruchinin_2013_PRB_87_115201}
S.{\relax Yu}. Kruchinin, M.~Korbman, V.S. Yakovlev, Phys. Rev. B \textbf{87},
  115201 (2013)

\bibitem{Story_1994_PRA_49_3875}
J.G. Story, D.I. Duncan, T.F. Gallagher, Phys. Rev. A \textbf{49}, 3875 (1994)

\bibitem{Paulus_2001_PRA_64_021401}
G.G. Paulus, F.~Grasbon, H.~Walther, R.~Kopold, W.~Becker, Phys. Rev. A
  \textbf{64}, 021401 (2001)

\bibitem{Kopold_2002_JPB_35_217}
R.~Kopold, W.~Becker, M.~Kleber, G.G. Paulus, J. Phys. B \textbf{35}(2), 217
  (2002)

\bibitem{Ivanov_CP_2013_414_3}
M.~Ivanov, O.~Smirnova, Chem. Phys. \textbf{414}(0), 3 (2013)

\bibitem{Minasian_1986_PRB_34_963}
H.~Minasian, S.~Avetisian, Phys. Rev. B \textbf{34}, 963 (1986)

\bibitem{Kovarskii_1971_PSSB_45_47}
V.A. Kovarskii, E.{\relax Yu}. Perlin, Phys. Stat. Sol. B \textbf{45}, 47
  (1971)

\bibitem{Gruzdev_2007_PRB_75_205106}
V.E. Gruzdev, Phys. Rev. B \textbf{75}, 205106 (2007)

\bibitem{Wirth_Science_2011}
A.~Wirth, M.T. Hassan, I.~Grguras, J.~Gagnon, A.~Moulet, T.T. Luu, S.~Pabst,
  R.~Santra, Z.A. Alahmed, A.M. Azzeer, V.S. Yakovlev, V.~Pervak, F.~Krausz,
  E.~Goulielmakis, Science \textbf{334}(6053), 195 (2011)

\bibitem{Yablonovitch_1989_PRL_63_976}
E.~Yablonovitch, J.P. Heritage, D.E. Aspnes, Y.~Yafet, Phys. Rev. Lett.
  \textbf{63}, 976 (1989)

\bibitem{Kuznetsov_1993_PRB_48_10828}
A.V. Kuznetsov, C.J. Stanton, Phys. Rev. B \textbf{48}, 10828 (1993)

\bibitem{Gertsvolf_2010_JPB_43_131002}
M.~Gertsvolf, M.~Spanner, D.M. Rayner, P.B. Corkum, J. Phys. B: At. Mol. Opt.
  Phys. \textbf{43}(13), 131002 (2010)

\bibitem{Mitrofanov_2011_PRL_106_147401}
A.V. Mitrofanov, A.J. Verhoef, E.E. Serebryannikov, J.~Lumeau, L.~Glebov, A.M.
  Zheltikov, A.~Baltu\ifmmode~\check{s}\else \v{s}\fi{}ka, Phys. Rev. Lett.
  \textbf{106}, 147401 (2011)

\bibitem{Wachter_PRL_2014}
G.~Wachter, C.~Lemell, J.~Burgd\"orfer, S.A. Sato, X.M. Tong, K.~Yabana, Phys.
  Rev. Lett. \textbf{113}, 087401 (2014)

\bibitem{Foldi_2013_NJP_15_063019}
P.~F\"oldi, M.G. Benedict, V.S. Yakovlev, New J. of Phys. \textbf{15}(6),
  063019 (2013)

\bibitem{Fortier_2004_PRL_92_147403}
T.M. Fortier, P.A. Roos, D.J. Jones, S.T. Cundiff, R.D.R. Bhat, J.E. Sipe,
  Phys. Rev. Lett. \textbf{92}, 147403 (2004)

\bibitem{Kurizki_1989_PRB_39_3435}
G.~Kurizki, M.~Shapiro, P.~Brumer, Phys. Rev. B \textbf{39}, 3435 (1989)

\bibitem{Atanasov_1996_PRL_76_1703}
R.~Atanasov, A.~Hach\'e, J.L.P. Hughes, H.M. van Driel, J.E. Sipe, Phys. Rev.
  Lett. \textbf{76}, 1703 (1996)

\bibitem{Hache_1997_PRL_78_306}
A.~Hach\'e, Y.~Kostoulas, R.~Atanasov, J.L.P. Hughes, J.E. Sipe, H.M. van
  Driel, Phys. Rev. Lett. \textbf{78}, 306 (1997)

\bibitem{Fraser_1999_PB_272_1999}
J.M. Fraser, A.I. Shkrebtii, J.E. Sipe, H.M. van Driel, Physica B
  \textbf{272}(1-4), 353  (1999)

\bibitem{Zhao_2008_JAP_103_053510}
H.~Zhao, E.J. Loren, A.L. Smirl, H.M. van Driel, J. Appl. Phys.
  \textbf{103}(5), 053510 (2008)

\bibitem{Costa_2007_NP_3_632}
L.~Costa, M.~Betz, M.~Spasenovic, A.D. Bristow, H.M. van Driel, Nature Physics
  \textbf{3}(9), 632 (2007)

\bibitem{Rioux_2012_PE_45_1}
J.~Rioux, J.E. Sipe, Physica E \textbf{45}(0), 1 (2012)

\bibitem{Franco_2007_PRL_99_126802}
I.~Franco, M.~Shapiro, P.~Brumer, Phys. Rev. Lett. \textbf{99}, 126802 (2007)

\bibitem{Durach_2010_PRL_105_086803}
M.~Durach, A.~Rusina, M.F. Kling, M.I. Stockman, Phys. Rev. Lett. \textbf{105},
  086803 (2010)

\bibitem{Durach_2011_PRL_107_086602}
M.~Durach, A.~Rusina, M.F. Kling, M.I. Stockman, Phys. Rev. Lett. \textbf{107},
  086602 (2011)

\bibitem{Apalkov_PRB_2012_86_165118}
V.~Apalkov, M.I. Stockman, Phys. Rev. B \textbf{86}, 165118 (2012)

\bibitem{Schiffrin_2014_Nature}
A.~Schiffrin, T.~Paasch-Colberg, N.~Karpowicz, V.~Apalkov, D.~Gerster,
  S.~Muhlbrandt, M.~Korbman, J.~Reichert, M.~Schultze, S.~Holzner, J.V. Barth,
  R.~Kienberger, R.~Ernstorfer, V.S. Yakovlev, M.I. Stockman, F.~Krausz, Nature
  \textbf{507}(7492), 386 (2014)

\bibitem{PaaschColberg_2014_NP_8_214}
T.~Paasch-Colberg, A.~Schiffrin, N.~Karpowicz, S.~Kruchinin, O.~Sa\u{g}lam,
  S.~Keiber, O.~Razskazovskaya, S.~M\"uhlbrandt, A.~Alnaser, M.~K\"ubel,
  V.~Apalkov, D.~Gerster, J.~Reichert, T.~Wittmann, J.V. Barth, M.I. Stockman,
  R.~Ernstorfer, V.S. Yakovlev, R.~Kienberger, F.~Krausz, Nature Photonics
  \textbf{8}(3), 214 (2014)

\end{thebibliography}

\end{document}